\documentclass[a4paper,fleqn,usenatbib,useAMS]{mnras}

\usepackage{amssymb}	
\usepackage{multicol}        
\usepackage{bm}		
\usepackage{pdflscape}	
\usepackage[T1]{fontenc}
\usepackage{ae,aecompl}

\usepackage {graphicx}
\usepackage{natbib}
\usepackage{aas_journals}
\usepackage{color}
\usepackage{amsmath}
\definecolor{mygray}{gray}{0.6}

\def\aap{{ A\&A}}

\def\apjl{{ApJL}}

\def\apjs{{ApJS}}

\def\redmagic{redMaGiC}

\newcommand{\bes}{\begin{equation*}}
\newcommand{\ees}{\end{equation*}}
\newcommand{\bea}{\begin{eqnarray}}
\newcommand{\eea}{\end{eqnarray}}
\newcommand{\beas}{\begin{eqnarray*}}
\newcommand{\eeas}{\end{eqnarray*}}

\newcommand{\mpc}{\rm {h^{-1}Mpc }}

\newcommand{\ltsima}{$\; \buildrel < \over \sim \;$}
\newcommand{\lsim}{\lower.5ex\hbox{\ltsima}}
\newcommand{\gtsima}{$\; \buildrel > \over \sim \;$}
\newcommand{\gsim}{\lower.5ex\hbox{\gtsima}}

\def\gtrsim{\mathrel{\hbox{\rlap{\hbox{\lower4pt\hbox{$\sim$}}}\hbox{$>$}}}}
\let\ga=\gtrsim
\def\lesssim{\mathrel{\hbox{\rlap{\hbox{\lower4pt\hbox{$\sim$}}}\hbox{$<$}}}}
\let\la=\lesssim

\newcommand{\be}{\begin{equation}}
\newcommand{\ee}{\end{equation}}
\newcommand{\ba}{\begin{eqnarray}}
\newcommand{\ea}{\end{eqnarray}}

\title[Cold ISW imprint of DES Y3 supervoids]{More out of less: an excess integrated Sachs-Wolfe signal from supervoids mapped out by the Dark Energy Survey}
\author[Andr\'{a}s Kov\'{a}cs et al.]{
\parbox{\textwidth}{
\LARGE
A.~Kov\'{a}cs$^{1,2}$\thanks{Corresponding author: \texttt{\rm \texttt{akovacs@ifae.es}}},
C.~S\'{a}nchez$^{3}$,
J.~Garc\'ia-Bellido$^{4}$,
J.~Elvin-Poole$^{5}$,
N.~Hamaus$^{6}$,
V.~Miranda$^{7}$,
S.~Nadathur$^{8}$,
T.~Abbott$^{9}$,
F.~B.~Abdalla$^{10}$,
J.~Annis$^{11}$,
S.~Avila$^{8}$,
E.~Bertin$^{12,13}$,
D.~Brooks$^{10}$,
D.~L.~Burke$^{14,15}$,
A.~Carnero Rosell$^{16,17}$,
M.~Carrasco~Kind$^{18,19}$,
J.~Carretero$^{1}$,
R.~Cawthon$^{20}$,
M.~Crocce$^{21}$,
C.~Cunha$^{14}$,
L.~N.~da Costa$^{17,22}$,
C.~Davis$^{14}$,
J.~De Vicente$^{16}$,
D.~DePoy$^{23}$,
S.~Desai$^{24}$,
H.~T.~Diehl$^{11}$,
P.~Doel$^{10}$,
E.~Fernandez$^{1}$,
B.~Flaugher$^{11}$,
P.~Fosalba$^{21}$,
J.~Frieman$^{11}$,
E.~Gazta\~{n}aga$^{21}$,
D. ~Gerdes$^{25}$,
R.~Gruendl$^{18,19}$,
G.~Gutierrez$^{11}$,
W.~Hartley$^{9,26}$,
D.~L.~Hollowood$^{27}$,
K.~Honscheid$^{28,29}$,
B.~Hoyle$^{30,6}$,
D.~J.~James$^{31}$,
E.~Krause$^{7}$,
K.~Kuehn$^{32}$,
N.~Kuropatkin$^{11}$,
O.~Lahav$^{10}$,
M.~Lima$^{33,17}$,
M.~Maia$^{17,22}$,
M.~March$^{3}$,
J.~Marshall$^{23}$,
P.~Melchior$^{34}$,
F. ~Menanteau$^{18,19}$,
C.~J.~Miller$^{25,35}$,
R.~Miquel$^{1, 35}$,
J.~Mohr$^{37,6,30}$,
A.~A.~Plazas$^{38}$,
K.~Romer$^{39}$,
E.~Rykoff$^{14,15}$,
E.~Sanchez$^{16}$,
V.~Scarpine$^{11}$,
R.~Schindler$^{15}$,
M.~Schubnell$^{25}$,
I.~Sevilla-Noarbe$^{16}$,
M.~ Smith$^{40}$,
R.~C.~ Smith$^{9}$,
M.~ Soares-Santos$^{41}$,
F.~Sobreira$^{42,17}$,
E.~Suchyta$^{43}$,
M.~Swanson$^{19}$,
G.~Tarle$^{25}$,
D.~Thomas$^{8}$,
V.~Vikram$^{44}$,
J.~Weller$^{37,30,6}$}
  \vspace{0.07cm}\\~\\
\parbox{\textwidth}{\centering \textsc{\Large(The DES Collaboration)} \\ \centering \textit{Author affiliations are listed at the end of this paper}\\ }}

\begin{document}
\date{Submitted 2018}
\pagerange{\pageref{firstpage}--\pageref{lastpage}} \pubyear{2018}
\maketitle
\label{firstpage}
\begin{abstract}
The largest structures in the cosmic web probe the dynamical nature of dark energy through their integrated Sachs-Wolfe imprints. In the strength of the signal, typical cosmic voids have shown good consistency with expectation $A_{\rm ISW}=\Delta T^{\rm data} / \Delta T^{\rm theory}=1$, given the substantial cosmic variance. Discordantly, large-scale hills in the gravitational potential, or supervoids, have shown excess signals. In this study, we mapped out 87 new supervoids in the total 5000 deg$^2$ footprint of the \emph{Dark Energy Survey} at $0.2<z<0.9$ to probe these anomalous claims. We found an excess imprinted profile with $ A_{\rm ISW}\approx4.1\pm2.0$ amplitude. The combination with independent BOSS data reveals an ISW imprint of supervoids at the $3.3\sigma$ significance level with an enhanced $A_{\rm ISW}\approx5.2\pm1.6$ amplitude. The tension with $\Lambda$CDM predictions is equivalent to $2.6\sigma$ and remains unexplained.

\end{abstract}
\section{Introduction}

The apparent dominance of the obscure dark energy is a great puzzle in modern cosmology. Nevertheless, the concordance $\Lambda$-Cold Dark Matter ($\Lambda$CDM) cosmological model has shown a remarkable stability and flexibility against major probes like the cosmic microwave background (CMB), type Ia supernovae, baryonic acoustic oscillations, redshift-space distortion measurements, galaxy clustering, and gravitational lensing. 

A complementary probe of dark energy is the integrated Sachs-Wolfe effect \citep[ISW]{SachsWolfe} in the linear regime and the subdominant Rees-Sciama effect \citep[RS]{ReesSciama} on smaller scales. The late-time decay of large-scale gravitational potentials, due to the imbalance of structure growth and cosmic expansion, imprints tiny secondary anisotropies to the primary fluctuations of the CMB as photons traverse these potentials \citep[see e.g.][]{Fosalba2003,Fosalba2004}. The details of the measured effect may unravel the dynamical properties of dark energy through the precise way in which it \emph{stretches} the largest cosmic structures.

The ISW signal in the $\Lambda$CDM model, however, is too weak to be directly reconstructed in the sea of primordial CMB photons. Therefore, this important complementary probe is, at best, expected to remain moderately informative about dark energy dynamics. Yet, \cite{CrittendenTurok1996} showed that the ISW effect may be measured in cross-correlations with tracers of the matter distribution with maximum signal-to-noise ratio $S/N\approx7.6$ for an idealistic deep full-sky survey. Practically, the expected significance remains at the $2<S/N<3$ level for currently available data sets (see e.g. \cite{Cabre2007}).

A combination of several tracer catalogues resulted in a constraint $A_{\rm ISW}=\Delta T^{\rm ISW}_{\rm data} / \Delta T^{\rm ISW}_{\rm \Lambda CDM}\approx1.00\pm0.25$ on the ISW ``amplitude" using angular cross-correlation techniques, where $A_{\rm ISW}=1$ corresponds to the concordance $\Lambda$CDM prediction \citep{gian, PlanckISW2015}. These combined measurements, despite their moderate signal-to-noise ratio, appear to be important consistency tests of alternative cosmologies. For instance, various Galileon models that predict a different sign for the ISW signal have practically been ruled out  \citep{Barreira2014,Renk2017}.

As an alternative, large voids and superclusters offer a way to reconstruct the ISW signal \emph{locally}. A pioneering measurement of this type by \cite{GranettEtAl2008} involves the identification of individual voids in the cosmic web using the \texttt{ZOBOV} algorithm \citep{ZOBOV}. Then CMB temperatures are stacked on the super-structure locations as a measure of their average imprint. The surprise was that the combined signal for supervoids and superclusters appears to be $\gsim3\sigma$ higher than $\Lambda$CDM expectations, according to theoretical and simulated follow-up studies \citep[e.g.][]{Nadathur2012,Aiola}. This curious signal has survived new CMB data releases and tests against systematics and remains a puzzle. 

Besides, voids also provide an interesting new window to cosmological observables in the low-density Universe, including baryonic acoustic oscillations \citep[e.g.][]{Kitaura2016}, Alcock-Paczy{\'n}ski tests \citep[e.g.][]{Sutter2012}, redshift-space distortions \citep[e.g.][]{Hamaus2016,Hawken2017}, or gravitational lensing \citep[e.g.][]{Melchior2014,Sanchez2016}. Synergies of the ISW measurements with these additional probes may uncover important new details about the apparent tensions in the amplitude of the signal.

\subsection{Voids vs. supervoids}

Naturally, revisions of the methods, tests of selection effects, and a possible confirmation in other data sets were crucial steps to (in)validate this apparent anomaly of the dark sector. Measurement at lower redshifts ($z<0.4$) using new Sloan Digital Sky Survey (SDSS) spectroscopic data showed no high-significance detection with differently constructed void catalogues \citep{Ilic2013,Planck19,CaiEtAl2014,Hotchkiss2015,Kovacs2015}. Ultimately, using the theoretically best possible stacking methods, simulations, and data available, \cite{NadathurCrittenden2016} recently reported a $3.1\sigma$ detection of the ISW signal from ``isolated" voids and superclusters in the Baryon Oscillations Spectroscopic Survey (BOSS) data release 12 (DR12). They used optimal matched filters and found $A_{\rm ISW}\approx1.65\pm0.53$, i.e. close to the most accurate estimates from full cross-correlations with $A_{\rm ISW}\approx1$ (see more detailed comparisons to our methods below).

Naively, these findings might sound conclusive but the situation is more intricate. As a clue for the special nature of the original SDSS supervoids, \cite{Granett2015} reconstructed their average shape using the overlapping BOSS DR12 spectroscopic data, and found that the supervoids are significantly \emph{elongated} in the line-of-sight. In photometric data, used also by \cite{GranettEtAl2008}, finding typical voids surrounded by overdensitites is challenging because of the smearing effect of photo-$z$ errors in the line-of-sight distribution of galaxies. Systems of ``merged" voids lined up in our line-of-sight constituting supervoids with numerous sub-voids, however, are possible to detect. Undeniably, elongated void structures have a longer photon travel time compared to a spherical void of the same angular size and therefore correspond to larger ISW temperature shifts \citep{KovacsJGB2015,MarcosCaballero2015}. Although this in principle could explain an excess signal, \cite{Flender2013} concluded that the assumption of sphericity does not lead to a significant underestimate of the ISW signal in a $\Lambda$CDM model. 

Relatedly, it is worth noting that \cite{Cai2016} did find excess signals using the BOSS DR12 data. They also focussed on efficient pruning strategies to, above all, remove the so-called voids-in-clouds that are expected to be aligned with hot spots on the CMB. Apart from the different filtering methods applied, most importantly \cite{Cai2016} also considered merged voids, while the implementation of the watershed algorithm by \cite{NadathurCrittenden2016} prevented neighboring voids from merging \citep[see also][]{NadathurEtal2016}. At least in part, this difference explains the different outcomes because \cite{Hotchkiss2015} have pointed out in simulations that the shape of the stacked ISW imprint does depend on the void definition. In particular, following \cite{GranettEtAl2008} to focus on the most extreme structures, \cite{Cai2016} only used voids with a \texttt{ZOBOV} probability measure $p_{\rm void}>3\sigma$ (i.e., least likely to occur in random data), and reported $A_{\rm ISW}\approx20$ at $3.4\sigma$ significance. This excess signal of large voids again suggests that void definition, and, in particular, details in the merging process of voids do have an important role in this problem. 

Then recently, \cite{Kovacs2017} critically revisited the above pruning and stacking strategies. Detailed simulation analyses validated the (dis)advantages of both strategies, and, importantly, proved that there is a rather special sub-population of large voids that leave a characteristic ``cold-spot-and-hot-ring" ISW profile with fine details. These extended underdensities of effective radii $R_{\rm v}\gsim100$ $\mpc$ encompass at least five merged sub-voids. In hindsight, these facts explain why \cite{Hernandez2013} found that varying the number of the objects in the stacking, or using different filter sizes lowers the significance, because the biggest fluctuations are also the rarest.

\cite{Kovacs2017} then performed yet another BOSS DR12 stacking measurement restricted to these supervoids and reported an excess signal with $A_{\rm ISW}\approx9$ at the $\approx2.5\sigma$ significance level. These findings are not affected by a posteriori bias arguments because a special sample of supervoids can be selected for stacking measurements prior to looking at real-world data. These supervoids appear to imprint an excess ISW signal on the CMB but independent new measurements are needed to validate these results further \emph{elsewhere} on the sky.

In \cite{Kovacs2016}, we have recently attempted to probe these claims in the Southern hemisphere. We used the first year data (Y1) of the Dark Energy Survey \citep[DES,][]{DES} and identified 52 voids and 102 superclusters at redshifts $0.2 < z < 0.65$ using the void finder tool described in \cite{Sanchez2016}. The heart of that method is a restriction to 2D slices of galaxy data, and measurements of the projected density field around centers defined by minima in the corresponding smoothed density field. Similarly to the \cite{Granett2015} analysis, our tests revealed a significant mean line-of-sight elongation for the super-structures that is caused by the photo-$z$ uncertainties. All in all, we found a $\Delta T \approx -10 ~\mu K$ cold imprint of voids, formally with $A_{\rm ISW}\approx8\pm6$, that is $1.2\sigma$ higher than the imprint of such super-structures in the simulated $\Lambda$CDM universe. We also found $A_{\rm ISW}\approx8\pm5$ for superclusters. Therefore, in combination we constrained $A_{\rm ISW}\approx8\pm4$ with DES Y1 super-structures. These measurements, although hinting again at a large ISW amplitude, were indecisive because of the significant noise level. 

In this paper, we extend these measurements to the full 5000 deg$^2$ footprint of the Dark Energy Survey. We also extend the redshift range of the analysis to $0.2<z<0.9$ with our new data to probe these anomalous results in the biggest volume available. Finally, we attempt to combine our improved DES measurements with existing BOSS results in order to reduce the statistical uncertainties and possibly put tight constraints on the $A_{\rm ISW}$ amplitude of supervoids.

The paper is organized as follows. Data sets and detection algorithms are introduced in Section 2. Our simulated and observational results are presented in Section 3, while the final section contains a summary, discussion and interpretation of our findings.

\section{Data sets for the ISW analysis}

\subsection{Maps of CMB temperature}
We use the {\it Planck} Spectral Matching Independent Component Analysis (SMICA) CMB temperature map \citep{Planck_15} for our cross-correlations with void positions. The map was downgraded to $N_{side}=512$ resolution (approximately $\sim6.87$ arcmin) with \texttt{HEALPix} pixelization \citep{healpix}. We masked out contaminated pixels with the $N_{side}=512$ WMAP 9-year extended temperature analysis mask \citep{WMAP9} to avoid re-pixelization effects of the $N_{side}=2048$ CMB masks provided by {\it Planck}. Several studies confirmed \citep[see e.g.][]{Planck19} that the cross-correlation signal observed at void locations is independent of the CMB data set when looking at WMAP Q, V, W, or {\it Planck} temperature maps. We, however, again checked for possible color dependence in the analysis.

\subsection{Galaxy data and mocks}

We closely follow \cite{Kovacs2016} in our methodology and extend our previous analysis by using photometric redshift data from the Dark Energy Survey \citep{DECam,morethanDE2016}. DES covers about one eighth of the sky (5000 deg$^2$) to a depth of $i_{AB} < 24$, imaging about 300 million galaxies in 5 broadband filters ($grizY$) up to redshift $z=1.4$. In this paper we used a luminous red galaxy sample from the first three years of observations (Y3). This Red-sequence MAtched-filter Galaxy Catalog \cite[\redmagic,][]{Rozo2015} is a catalog of photometrically selected luminous red galaxies, based on the red-sequence MAtched-filter Probabilistic Percolation (redMaPPer) cluster finder algorithm \citep{Rykoff2014}. We utilized the \redmagic\ sample because of its exquisite photometric redshifts, namely $\sigma_z/(1+z)\approx 0.02$, and a $4\sigma$ redshift outlier rate of $r_\mathrm{out}\simeq1.41\%$. The resulting galaxy sample has a constant comoving space density in three versions, $\bar{n}\approx 10^{-3}h^{3}$ Mpc$^{-3}$ (high density sample, brighter than 0.5$L_{*}$), $\bar{n}\approx4\times10^{-4}h^{3}$ Mpc$^{-3}$ (high luminosity sample, brighter than 1.0$L_{*}$)), $\bar{n}\approx1\times10^{-4}h^{3}$ Mpc$^{-3}$ (higher luminosity sample, brighter than 1.5$L_{*}$)). 

In general, we aim to choose the free parameters of our analysis and prune our void catalogue in order to detect and then study supervoids rather than ordinary voids defined with other methods. Different void definitions are optimal for different cosmological probes as shown by \cite{Cautun2018} in the case of void lensing signals. For ISW analyses using voids, \cite{Kovacs2017} showed that supervoids of radii $R_{\rm v}\gsim100$ $\mpc$ represent a special subclass of extended underdensitites with a specific ISW imprint. Importantly, these large voids have shown anomalies previously that motivate further studies with more details using DES data.

Therefore, we decided to use the higher luminosity tracer sample because it has an approximately constant co-moving space density of tracers up to $z_{\rm max}=0.9$ (as opposed to $z_{\rm max}=0.65$ and $z_{\rm max}=0.8$ for the other versions) that is a key property to maximize the volume of our supervoid mapping at the expense of surveying the volume of interest with a sparser sample. In fact, in sparser galaxy tracers the number of voids identified decreases, but the average void size is larger \citep{Sutter2014b}. More importantly, voids resolved by sparse galaxy samples are also on average shallower but trace more extended dark matter underdensities (or supervoids) which should have a longer photon travel time and therefore correspond to larger ISW temperature shift \citep[see e.g.][]{Hotchkiss2015,Nadathur2015}. 

For reconstructing the $\Lambda$CDM expectations of the stacked ISW imprint of DES Y3 supervoids, we closely followed our methodology developed for DES Y1 \citep[see][]{Kovacs2016}. We used the full-sky simulated ray-tracing temperature data from the Jubilee ISW project \citep{Watson2014}. This project is built upon the Jubilee simulation, a $\Lambda$CDM N-body simulation with $6000^3$ particles in a volume of ($6h^{-1}$ Gpc)$^{3}$, assuming WMAP-5 cosmology. A corresponding full-sky mock LRG catalogue was initially designed to model the properties of SDSS LRGs studied in \cite{Eisenstein2005}. This mock provides a sample with $\bar{n}\approx8\times10^{-5}h^{3}$ Mpc$^{-3}$ that is comparable to the galaxy density of our DES higher luminosity \redmagic\ sample. The Jubilee mock also features a constant comoving space density like the DES higher luminosity sample.

We note that void finding is not only sensitive to sparsity, but also more highly biased tracers yield larger voids on average \citep{Sutter2014a,Pollina2017}. Recently, \cite{Pollina2018} studied the roles of tracer bias and photo-$z$ errors in void properties using DES mocks and observations. They also reported a larger mean void size when using sparser tracers. Importantly, they also confirmed that large voids can be robustly identified using DES photo-$z$ data focusing on galaxy clusters.

Nevertheless, the estimated linear galaxy bias values of the Jubilee and DES LRG samples are quite similar at the level of $b_{g}\approx2.0$, thus no meaningful difference is expected in this respect \citep{ho, Elvin-Poole2017}. \cite{Hotchkiss2015} and \cite{Flender2013} both found that the expected stacked ISW signal one determines from Jubilee, or from similar ISW simulations, will always be an \emph{overestimate} of that observable from superstructures in DES-like data, especially if the tracer density in the mock is lower. Relatedly, \cite{Hernandez2013} reported, in yet another ISW simulation and modeling analysis, that Gaussian and fully non-linear simulations are in close agreement. The difference induced by adopting a slightly different cosmological model should introduce changes in the ISW amplitude at the 2 per cent level. We conclude, therefore, that the simulated Jubilee results can meaningfully be compared with observed DES data to estimate the level of (in)consistency between the ISW imprint of supervoids.

\begin{table}
\centering
\caption{Number of voids (supervoids) in the Jubilee mock and corresponding central ISW temperatures in the stacked samples as a function of density smoothing. We note that, as expected, larger smoothing values result in a monotonously lower number of voids in general. However, as a result of larger smoothing and merging of smaller voids into fewer larger ones where possible, the relative number of supervoids of radii $R_{v} > 100$ Mpc/h in the sample increases. Consequently, the average size of the less numerous supervoids increases for larger smoothings, corresponding to colder central ISW imprints but with fewer objects to stack. The $50~h^{-1}{\rm Mpc}$ choice guarantees a preferable combination of the highest number of supervoids with relatively high expected signal amplitude.}
\begin{tabular}{@{}ccc}
\hline
\hline
Smoothing & $N_{v}$ ($R_{v} > 100$ Mpc/h) & $\Delta T (R/R_{v}=0.1)$  \\
\hline
20 $Mpc/h$ & 4538 (1083) & -1.8 $\mu K$\\
\hline
50 $Mpc/h$ & 2019 (1793) & -2.2 $\mu K$\\
\hline
100 $Mpc/h$ & 272 (248) & -3.3 $\mu K$\\
\hline
\hline
\end{tabular}
\label{table:merging}
\end{table}

\subsection{Supervoids for stacking}

We now describe our methodology for void definition and necessary pruning techniques. In a DES void finding project, \cite{Sanchez2016} found that significant real underdensities can be identified even using photo-$z$ data in tomographic slices of width roughly twice the typical photo-$z$ uncertainty. The low outlier rate of the DES \redmagic\ sample makes Gaussian photo-$z$ errors with $\sigma_z/(1+z)\approx 0.02$ added to Jubilee redshifts sufficient to model the \redmagic\ characteristics for our purposes.

The heart of the method is a restriction to 2D slices of galaxy data, and measurements of the projected density field around centers defined by minima in the corresponding smoothed density field. The line-of-sight slicing was found to be appropriate for slices of thickness $2s_{v}\approx100~h^{-1}{\rm Mpc}$ for photo-$z$ errors at the level of $\sigma_{z}/(1+z) \approx 0.02$ or $\sim50~h^{-1}{\rm Mpc}$ at $z\approx0.5$. We adopt this strategy in our Jubilee void finding procedure and slice the data in shells of $100~h^{-1}{\rm Mpc}$ thickness in the line-of-sight.

A free parameter in the method is the scale of the initial Gaussian smoothing applied to the galaxy density field. In \cite{Kovacs2016}, we have found, in the Jubilee simulation, that $\sigma=20~h^{-1}{\rm Mpc}$ is a preferable choice for ISW measurements using the whole void sample in the stacking procedure. For weak lensing measurements with DES voids, however, \cite{Sanchez2016} reported that the smaller $\sigma=10~h^{-1}{\rm Mpc}$ smoothing is preferable. In order to more efficiently identify the largest structures with possible merging of smaller voids, we now increased the smoothing length of the void finder. With larger smoothing, the merging of typical voids into larger encompassing supervoids becomes possible. 

While a $\sigma=20~h^{-1}{\rm Mpc}$ smoothing results in more voids in total, as a result of void merging, a $\sigma=50~h^{-1}{\rm Mpc}$ smoothing returns significantly more $R_{\rm v}\gsim100$ $\mpc$ supervoids that we aim to study. Moreover, we found that a $\sigma=50~h^{-1}{\rm Mpc}$ smoothing results in a $\sim25\%$ higher signal than with $\sigma=20~h^{-1}{\rm Mpc}$ thus we certainly gain in terms of signal-to-noise. We summarize our related analyses in Table 1.

We note that a significantly larger smoothing scale is not expected to further increase the detectable signal. Firstly, the number of super-structures would decrease significantly, even if the signal amplitude is slightly increased compared to the $\sigma=50~h^{-1}{\rm Mpc}$ smoothing case. Furthermore, the smoothing level we chose is comparable to the photo-$z$ smearing effect in the line-of-sight direction, resulting in fairly spherical overall smoothing without significant elongation in our line-of-sight. We thus adopt a $\sigma=50~h^{-1}{\rm Mpc}$ smoothing scale in our analysis to further optimize the measurement for supervoids.

Another arbitrary choice in the analysis is the definition of the slice boundaries along the line-of-sight. Following again \cite{Kovacs2016}, we created supervoid catalogues using shifted ``slicings" of the galaxy catalogue for both data and simulations and tested the consistency among the different resulting catalogues in terms of general catalogue properties and measurement characteristics. In our measurement, we consider the mean signal coming from these slightly different realizations of the slicing.

\section{Results}

\subsection{Simulations - a Jubilee analysis}

We build up on the recent findings by \cite{Kovacs2017} who reported that the excess ISW signals, seen also by other authors with sub-optimal pruning techniques, can be attributed to supervoids of radii $R_{\rm v}\gsim100$ $\mpc$ (for details in void definition and methodology see \cite{Sanchez2016}). Importantly, the stacked ISW imprint of these large-scale features in the density map shows a cold spot surrounded by a hot ring, unlike for smaller voids. This suggests that this is a physically different branch of underdensities that efficiently probe the largest hills of the gravitational potential in their full extent. These super-structures actually carry most of the potentially observable ISW signal which, while environment, density profiles, redshifts, and exact shapes can be important for the accurate estimates, certainly is expected to correlate with void size. 

Nevertheless, if the SDSS/BOSS excess ISW-like signal of these supervoids is simply a pattern in noise, it certainly should not occur \emph{elsewhere} in the sky. Therefore, the independent DES data is a great way to test these claims. We thus follow up on these anomalous results and \emph{a priori} consider these seemingly anomalous $R_{\rm v}\gsim100$ $\mpc$ fluctuations in the gravitational potential to possibly confirm or falsify the apparent excess signals.

We re-estimated the expected ISW imprint of these supervoids using the Jubilee lightcone. We identified $1793$ mock supervoids at redshifts $0.2<z<0.9$, providing a basis for accurate calculation of the expected signal. Following \cite{Kovacs2016}, we stack the ISW-{\it only} Jubilee temperature map on supervoid locations using the \texttt{gnomview} projection technique of \texttt{HEALPix} \citep{healpix}. We also re-scale the images knowing the angular size of the supervoids thus the void boundaries coincide in the stacked image as well as the centres. 

\begin{figure}
\begin{center}
\includegraphics[width=90mm]{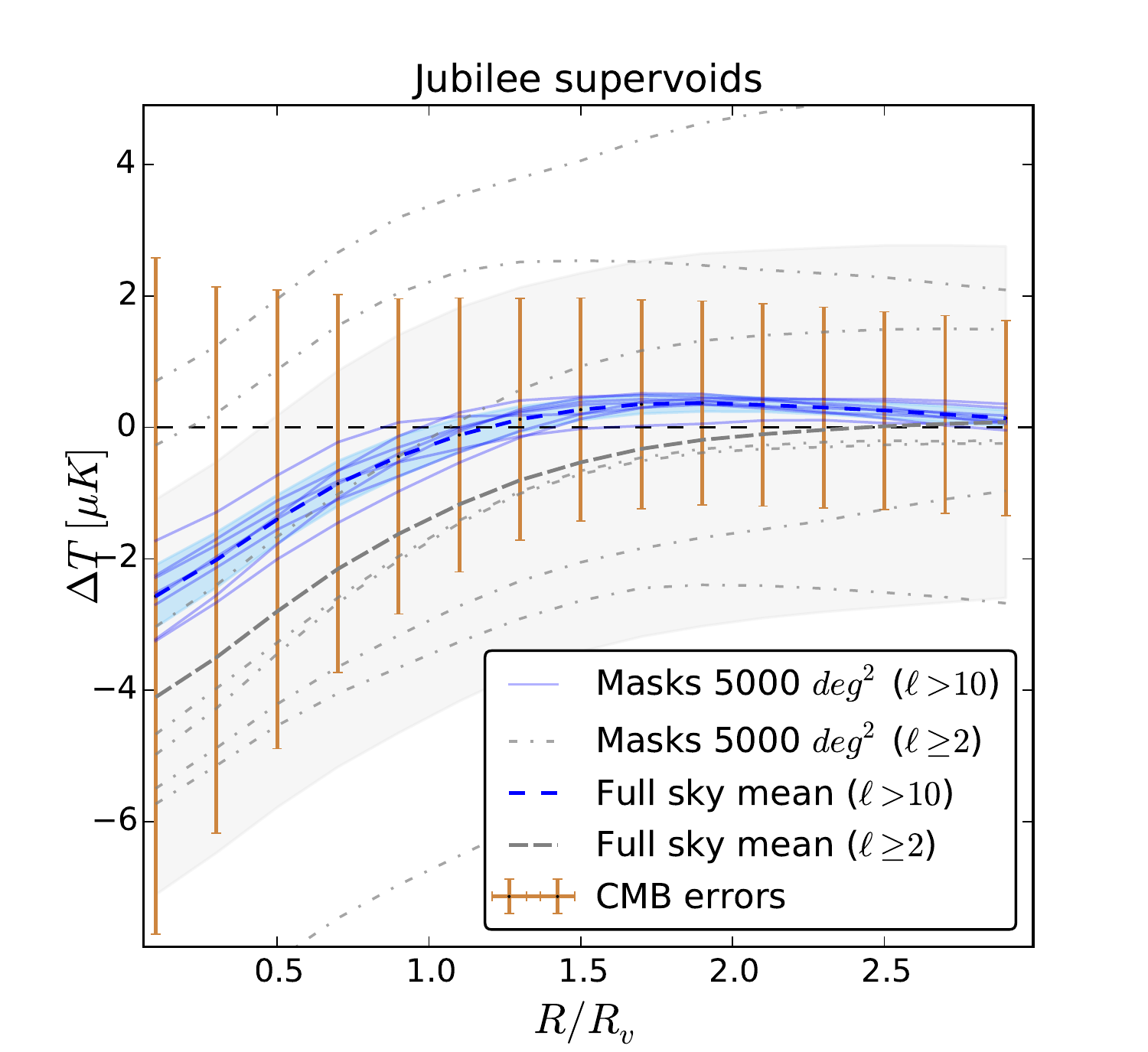}
\caption{Stacked ISW profiles of supervoids are compared with and without $2\leq \ell \leq10$ modes in the ISW map of the Jubilee simulation. Thick curves correspond to full sky estimates in Jubilee and shaded areas mark the sample variance when considering 8 different octants in Jubilee (the DES Y3 footprint is also of similar size with 5000 deg$^{2}$). Thin sets of curves show the individual stacked signals in different octants while error bars indicate the dominant CMB errors (around the $\ell>10$ full sky profile) for the DES Y3 window given the void catalogue properties in our data. Results using $\ell>10$ modes in the octants show less variation around the full sky result, while tests including $2\leq \ell \leq10$ modes show stronger variation and biases. These findings indicate that for a sufficient DES Y3 analysis with a masked footprint the removal of the large-scale modes is necessary for an unbiased estimate.}
\end{center}
\end{figure}

\begin{figure*}
\begin{center}
\includegraphics[width=175mm]{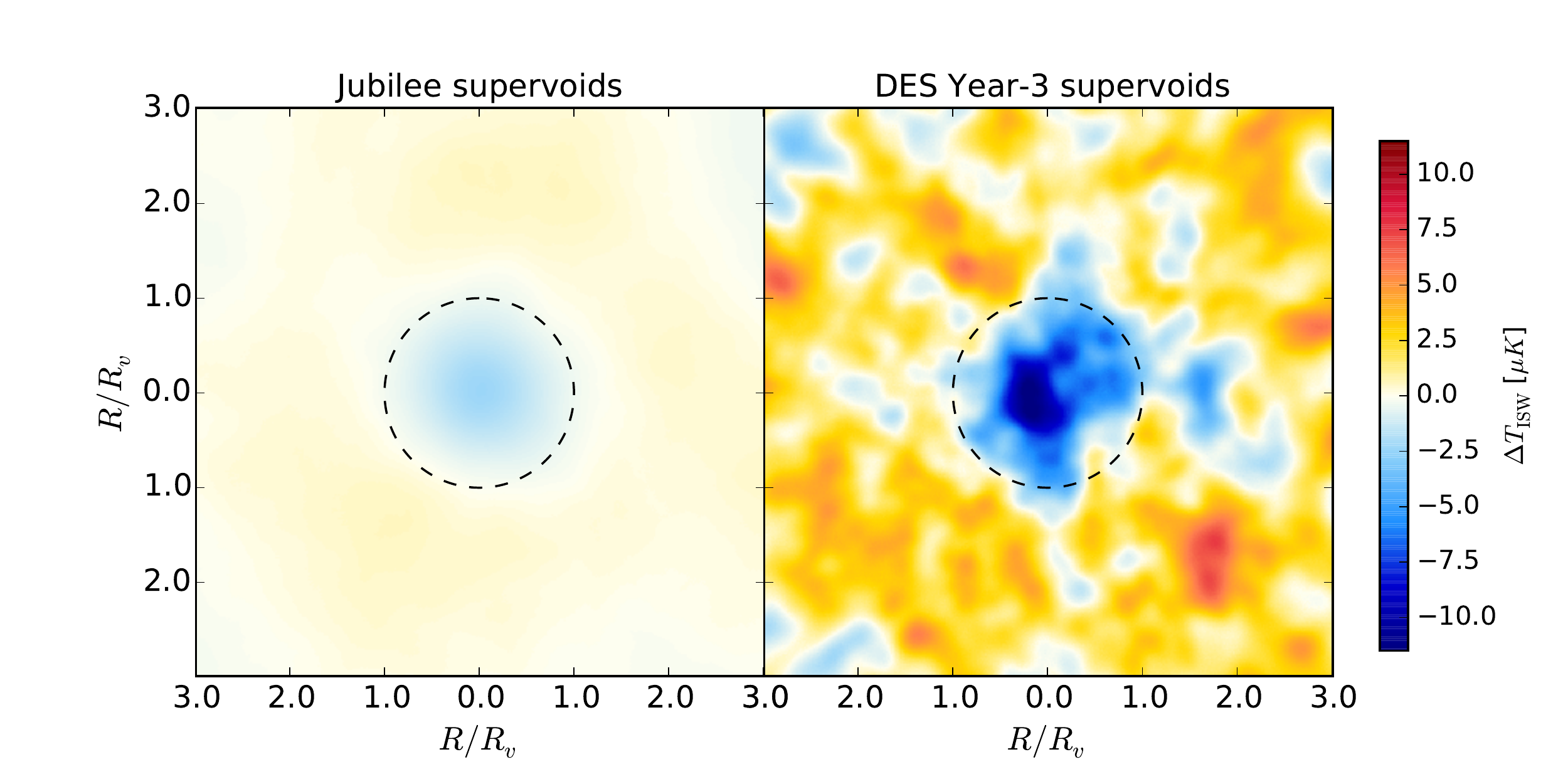}
\caption{Stacked ISW signals for Jubilee supervoids (left) and DES Year-3 supervoids (right). Dashed central circles mark the radius of the supervoids in re-scaled units. Results are presented using identical color scales. For DES data, we applied a smoothing to the individual raw CMB images only for this illustration using $\sigma=2^{\circ}$ symmetrical Gaussian beam in \texttt{HEALPix}.}
\end{center}
\end{figure*}

Given the measurement errors, \cite{Kovacs2017} concluded that the removal of the $2\leq \ell \leq10$ large-scale ISW modes helps to remove potential biases from the measured profiles, at the expense of reducing the signal itself by possibly a factor of two. The result is a more accurate mapping of features and a better convergence to zero signal at $R/R_{v}\approx2$ for these largest voids. We repeated these tests with our Jubilee supervoids and confirmed the previous results. As shown in Figure 1, the $2\leq \ell \leq10$ modes in Jubilee result in a convergence to zero signal is reached at higher radii far beyond the actual void radius. Relevantly, the treatment of these large-scale fluctuations is yet another difference between the different outcomes of two recent BOSS DR12 analyses; \cite{Cai2016} removed the $2\leq \ell \leq10$ modes while \cite{NadathurCrittenden2016} used all available modes, and only the former reported unexpected excess signals.

Another important element of the accurate estimation of the expected ISW signal is the role of masking. In fact, masking is related to the removal of the large-scale $2\leq \ell \leq10$ modes. If the 5000 deg$^{2}$ DES Y3 footprint happens to cover a hotter or colder large-scale peak in the observable ISW map, then such contributions from super-survey modes may bias the signal. This is a further argument in support of the removal of these $2\leq \ell \leq10$ modes.

In \cite{Kovacs2016} we showed that there is a non-negligible intrinsic fluctuation in the ISW imprint of supervoids if the signal is estimated from rather small DES Y1-like 1000 deg$^{2}$ cut-sky samples in Jubilee, especially if all modes are considered down to $\ell=2$. We now considered independent DES Y3-like 5000 deg$^{2}$ masked areas in eight octants on the Jubilee sky. In Figure 1, we demonstrate that an unbiased estimate of the ISW signal is obtained, with tolerable variation in the full extent of the imprint profile, if the large-scale modes are excluded from the analysis. With $2\leq \ell \leq10$ modes included, we find non-negligible bias and higher sample variance for the masked octant skies that we analyzed.

Naturally, considering the full-sky result the accuracy of the stacked profile can be further increased with more supervoids in the stacked sample. We therefore use the full-sky estimate in Jubilee for an accurate and unbiased estimate of the stacked ISW imprint of supervoids.

We note that, in principle, the larger area of the DES Y3 data compared to the Y1 subset may allow the consideration of lower modes (possibly $5\leq \ell \leq10$). However, we do not include them in our analysis as they correspond to angular sizes that exceed that of our supervoids ($3.5^{\circ} \lsim R_{\measuredangle} \lsim 15.9^{\circ}$). Moreover, the comparison of our results to earlier estimations of stacked ISW imprints, that typically removed $2\leq \ell \leq10$, becomes easier.

\subsection{The Dark Energy Survey data}

We simply repeat the above stacking procedure for a sample of 87 DES supervoids of radii $R_{\rm v}\gsim100$ $\mpc$. Reassuringly, we only removed 5 voids from the sample that were smaller than $100$ $\mpc$ thus the optimization of our void finding algorithm to form supervoids of voids if possible with large initial smoothing works with great efficiency. We also note that we robustly find $87\pm3$ similar supervoids if the ``slicing" of the density field is shifted by up to $40$ $\mpc$, accounting for the arbitrary choice of slice definition as explained in Section 2.3. The few biggest, though rather shallow, observed supervoids reach radii $R_{\rm v}\approx250$ $\mpc$. This is virtually compatible with the expected abundance of these largest, and typically rather shallow, structures in the matter density field \citep{Nadathur2014,SzapudiEtAl2014}.

The stacked CMB image of the DES supervoids is shown in Figure 2, together with the simulated result. While the images are similar in the nature of the imprints with cold spots in the center and a typically hot surrounding area, the amplitude of the signal is higher for the DES data. The data shows a visually compelling $\Delta T \approx -10 ~\mu K$ cold imprint in the central region of the DES image. For comparison, the coldest pixels in the Jubilee image are of $\Delta T \approx -2 ~\mu K$, in consistency with several previous results \citep{Nadathur2012,Flender2013,Hernandez2013}.

Importantly, the angular extent of the central DES cold spot appears to be consistent with the simulated result. We note that both \cite{Kovacs2016} and \cite{Kovacs2017} found, empirically, a $30\%$ mismatch in the angular size of the central cold spots for DES and BOSS supervoids, respectively, when compared to Jubilee results. At least in part, these mismatches are sourced by differences in tracer density in data and simulation in these previous studies. We interpret this better agreement between data and sims as a consequence of improvements that we implemented in the catalogue selection and pruning of DES data to match the Jubilee mock more accurately.

\begin{figure*}
\begin{center}
\includegraphics[width=173mm]{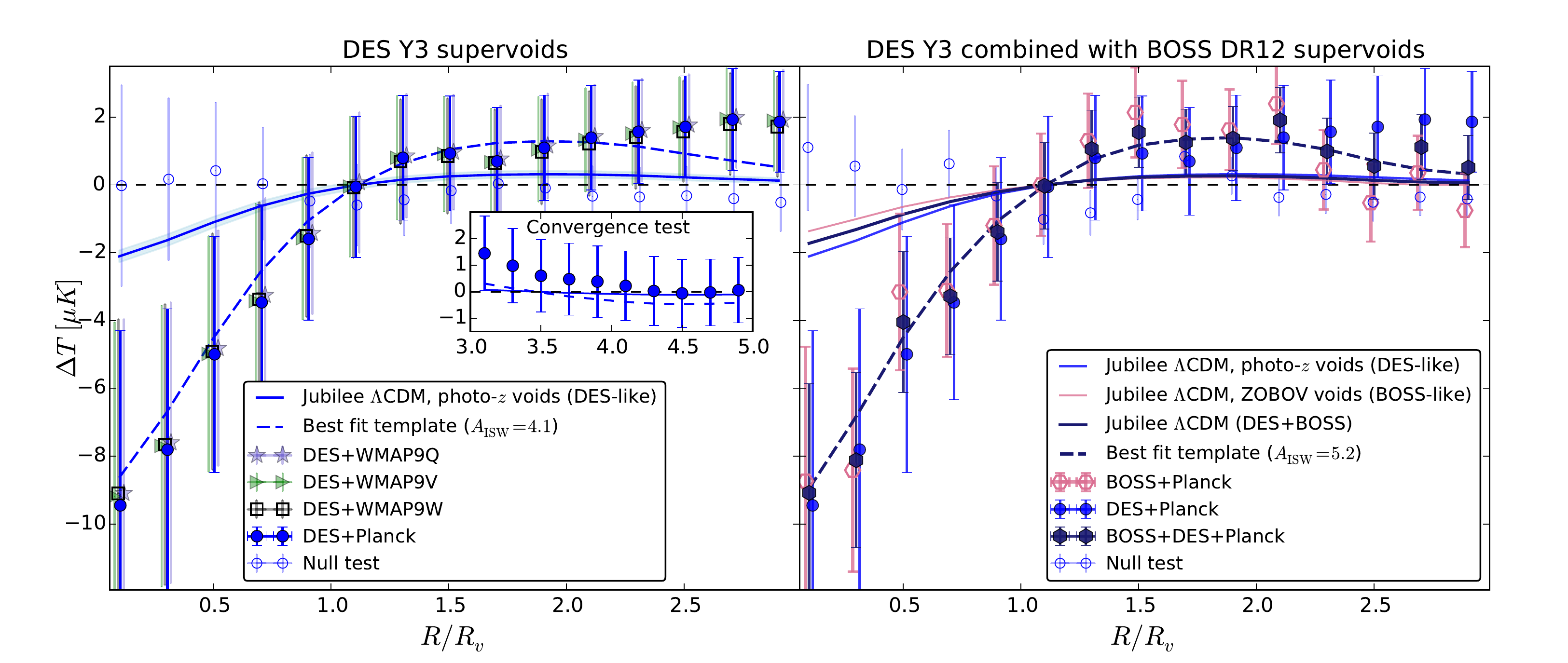}
\caption{{\it Left:} template fitting results of DES supervoids are compared to those of Jubilee supervoids (identified as in DES data). For completeness, we estimated the uncertainty of the simulated signal by random stackings in the Jubilee simulation (shaded region around the solid Jubilee curve). No CMB color dependence was observed. {\it Right:} A combination with BOSS supervoid results by Kov\'acs (2018). We found evidences for a rather high $A_{\rm ISW}$ amplitude for both DES and BOSS+DES. Note that the measured signals are quite similar in most of the profile, while the estimated signals are different for the DES and BOSS supervoid samples. Most probably, the elongated shape of the DES supervoids is responsible for this difference. The error bars are based on the $1000$ random stacking measurements using Gaussian CMB simulations that we describe in Section 3.3.}
\end{center}
\end{figure*}

\subsection{Template fitting: DES Y3 and BOSS DR12}

Beyond the visual impressions, we measured the azimuthally averaged radial ISW profile in $R/R_{v}$ fractional void radius units with bins of $\Delta (R/R_{v})=0.2$. We extend the range of the profile to $R/R_{v}=3$ to include the potentially measurable ISW imprints of the large-scale overdensities that surround the supervoids in the cosmic web. 

Overall, the DES Y3 data shows an ISW amplitude $A_{\rm ISW}\approx4.1\pm2.0$ with a moderate $2.1\sigma$ significance level using the {\it Planck} SMICA map. This translates to a $1.6\sigma$ excess signal above $A_{\rm ISW}=1$. As an additional test, we do not find evidence for significant frequency dependence when using WMAP9 Q, V, and W temperature maps (see left panel of Figure 3). Relatedly, small-scale systematic effects in the DES galaxy maps are not expected to affect our results, given on one hand the large angular size ($R_{\measuredangle} \gsim 3.5^{\circ}$) of the supervoids. On the other hand, the cross-correlation nature of our measurement also makes harder for the survey systematics to alter the observed correlations, because spurious voids would only dilute the stacked signal \citep[see e.g.][for related analyses and relevance in DES galaxy clustering probes]{Elvin-Poole2017}. We also performed null tests by rotating the {\it Planck} SMICA map around the Galactic poles by $\pm90^{\circ}$ and by randomizing the void positions in the DES window. We found no spurious correlations (see Figure 3). 

We also analyzed in greater details the convergence properties of the stacked profile in the outer profile at $R/R_{v}\approx3$. While the DES data points show a potential trend for a bias in our measurements at large radii, the continuation of the profile to $R/R_{v}\gsim3$ clearly demonstrate that, given the error bars, the profile converges to zero signal away from the void interior as expected (see the inset in the left panel of Figure 3). We thus argue that the highest measured temperature in the DES Y3 profile at $R/R_{v}\gsim2.7$ is a fluctuation given the error bars and it does not correspond to a real peak in the ISW imprint profile that is expected at $R/R_{v}\approx1.8$ based on our Jubilee analysis.

In our methodology, we fit an $A_{\rm ISW}$ amplitude to the observable imprints in the DES data using the Jubilee ISW template profile we constructed. We evaluated a statistic 
\begin{equation}
{\chi}^2 = \sum_{ij} (\Delta T_i^\rmn{DES}-A_{\rm ISW}\Delta T_i^\rmn{Jub} )C_{ij}^{-1} (\Delta T_j^\rmn{DES}-A_{\rm ISW}\Delta T_j^\rmn{Jub})
\end{equation}
where $C$ is the covariance matrix obtained by performing $1000$ random stacking measurements using Gaussian CMB simulations. The randoms have been generated with the \texttt{HEALPix} \texttt{synfast} routine using the {\it Planck} 2015 data release best fit CMB power spectrum \citep{Planck_15}. Gaussian CMB simulations without instrumental noise suffice because the CMB error is dominated by cosmic variance on the scales we consider \citep[see][]{Hotchkiss2015}. 

We first determined the sample variance associated with the DES Y3 window on the simulated CMB skies ($\ell>10$ modes included) as each masked random map has a different non-zero mean temperature that adds a bias to the stacked images. We found that the standard deviation of these fluctuations is $\sigma(\Delta \bar{T}_{DES})\approx1.1~\mu K$. We then found a fairly typical $\Delta \bar{T}_{DES}\approx0.3~\mu K$ bias value in the filtered {\it Planck} temperature map. We de-biased the observed temperature profile and each simulated CMB map in the masked DES Y3 window, and tested the effect of this correction on the resulting covariance matrices and errors. When removing the bias, we found a moderate $\approx 10\%$ larger noise inside the re-scaled void radius ($R/R_{v}<1$) and a rather important $\approx 50-60\%$ increment in the errors and stronger bin-to-bin covariance in the outer profile at $R/R_{v}\gsim2$ (see Figure 4 for a visual impression).

We then repeated the stacking procedure on the simulated CMB skies. A potential strategy to estimate the error bars is to keep the void positions fixed and vary the CMB realization, because in this case overlap-effects for voids are accounted for more efficiently \citep[see related discussions in][]{Hotchkiss2015}. We note that having overlapping supervoids does not automatically introduce a bias in the measurement because we estimate the signal in Jubilee with the same procedure, instead of modeling individual structures. Importantly, \cite{Flender2013} analyzed the differences between a spherical model of Gaussian perturbations and fully simulated ISW maps with ray-tracing. The latter contain contributions from potentially very elongated super-structures which add more to the total signal than only spherical structures. They reported that the differences are negligible thus overlapping voids that may form elongated structures in the line-of-sight do not significantly affect the results.

However, \cite{Cabre2007} showed in their simulated analyses of ISW error estimation methods in comparison that keeping a single realization of the galaxy map that one cross-correlates with the simulated CMB skies results in a $\approx10\%$ under-estimation of the true measurement errors. For DES Y3 \redmagic\ data, a large set of mock galaxy catalogues is not available (only five at the moment) to completely solve this problem but we performed two related tests to check how a stacking measurement using voids is affected. As an external test, we first considered 1000 mock BOSS supervoid catalogues\footnote{produced by Seshadri Nadathur, available on request} and the BOSS DR12 supervoid sample with 96 observed supervoids used by \cite{Kovacs2017}. We confirmed the general galaxy-CMB cross-correlation results by \cite{Cabre2007} for supervoid samples, finding $\approx9\%$ larger error bars (and slightly altered correlation structure) when varying both the CMB and the galaxy maps in the random stacking. 

We then cross-correlated the random CMB maps with randomly selected mock DES Y3 \redmagic\ void catalogues out of the five mocks available. The random runs, as expected, showed a null result for a correlation on average. We again determined $\approx9\%$ larger error bars compared to keeping the observed DES Y3 catalogue in front of all random CMB maps. These findings convincingly demonstrate that an approximately accurate correction can be implemented even with a smaller number of realistic mock supervoid catalogues. We thus include these corrections in our signal-to-noise analyses. The resulting covariance matrices are shown in Figure 4. We note that the shape of the $\chi^{2}$ histogram considering simulated fits of the $A_{\rm ISW}$ amplitude in randoms closely follows a $\chi^{2}$ distribution with 14 degrees of freedom (given 15 bins and one fit parameter) and the $A_{\rm ISW}$ values themselves closely approximate a Gaussian distribution with zero mean.

We then combined the DES Y3 measurements with BOSS DR12 supervoid results by \cite{Kovacs2017}. These supervoids in BOSS, and their corresponding BOSS-like Jubilee supervoids, are based on a different void finding algorithm called \texttt{ZOBOV} (ZOnes Bordering On Voidness)\footnote{The DR12 and Jubilee catalogues used by \cite{Kovacs2017} are non-public. However, a public catalogue is available for BOSS DR12 and DR11 at \texttt{http://www.icg.port.ac.uk/stable/nadathur/voids/} or on request from their authors \citep[see][for details]{Nadathur2016}.}. This widely used algorithm identifies density depressions in a 3-dimensional set of points and relies on watershed techniques to map the full extent of the under-densities \citep[see e.g.][for details]{ZOBOV,NadathurEtal2016}. 

\begin{figure}
\begin{center}
\includegraphics[width=78mm]{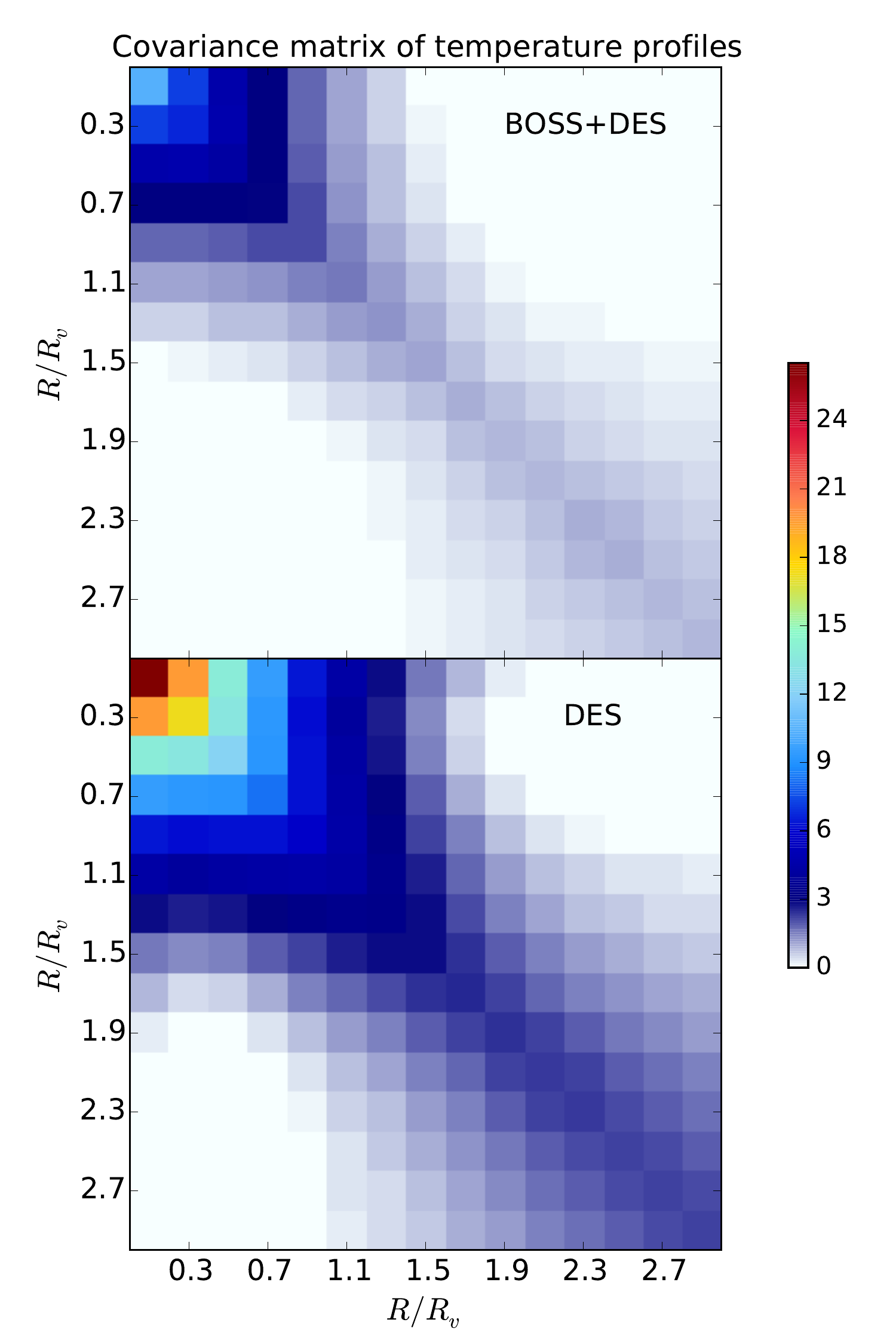}
\caption{Covariance matrices of the stacked temperature profiles for the DES Y3 data (bottom) and the BOSS+DES combined sample that we used in the template fitting analysis.}
\end{center}
\end{figure}

Overall, we report good consistency between the observed DES and BOSS data points with slight differences in the outer profiles beyond $R/R_{v}\gsim2$, as shown in the right panel of Figure 3. The joint DES+BOSS measurement shows a central cold spot and a surrounding hot ring. Quantitatively, the combination with 96 independent supervoids reveals an excess ISW signal of supervoids with $A_{\rm ISW}\approx5.2\pm1.6$ amplitude, corresponding to $3.3\sigma$ significance.

Notably, the $\Lambda$CDM expectation for the DES supervoids shows a colder imprint in the center while having comparable hot rings for BOSS and DES void populations. We interpret this finding as a consequence of the preference for supervoids moderately elongated in our line-of-sight due to photo-$z$ smearing, as discussed by \cite{Kovacs2016} and also in this paper in Section 1.1. However, the BOSS observations and BOSS-like \texttt{ZOBOV} voids in the Jubilee simulations have also shown similar features in the stacked signal thus we conclude that elongation itself is not the main source of these anomalies.

All things considered, this result underlines the importance of a better understanding of void finding strategies for ISW measurements. In Figure 4, we compare our constraints to several relevant observational probes of the amplitude of the ISW signal. By combining the independent results for supervoid observations, the discrepancy with concordance $\Lambda$CDM expectations grows to $2.6\sigma$ for BOSS+DES ($1.6\sigma$ for DES alone).

\section{Discussion \& Conclusions}

We measured the cold imprint of DES supervoids on the CMB using the full 5000 deg$^2$ survey area ($\sim1500$ deg$^2$ in our Year-1 analysis). We also extended the redshift range for supervoid search from $z\approx0.65$ to $z\approx0.9$. We argued that the expected signal of supervoids is higher when combining a large density smoothing and a sparser tracer sample (higher luminosity). With these \emph{a priori} analysis choices, our data is more similar to the Jubilee mock catalogues that we use to estimate the ISW signal. We mapped out 87 supervoids to probe previous claims of excess ISW signals coming from these vast hills imbedded in the gravitational potential. For DES data alone, we found an excess imprinted profile with $ A_{\rm ISW}\approx4.1\pm2.0$ amplitude. This observed value is higher than the Jubilee $\Lambda$CDM estimate with a moderate $1.6\sigma$ confidence.

\begin{figure*}
\begin{center}
\includegraphics[width=175mm]{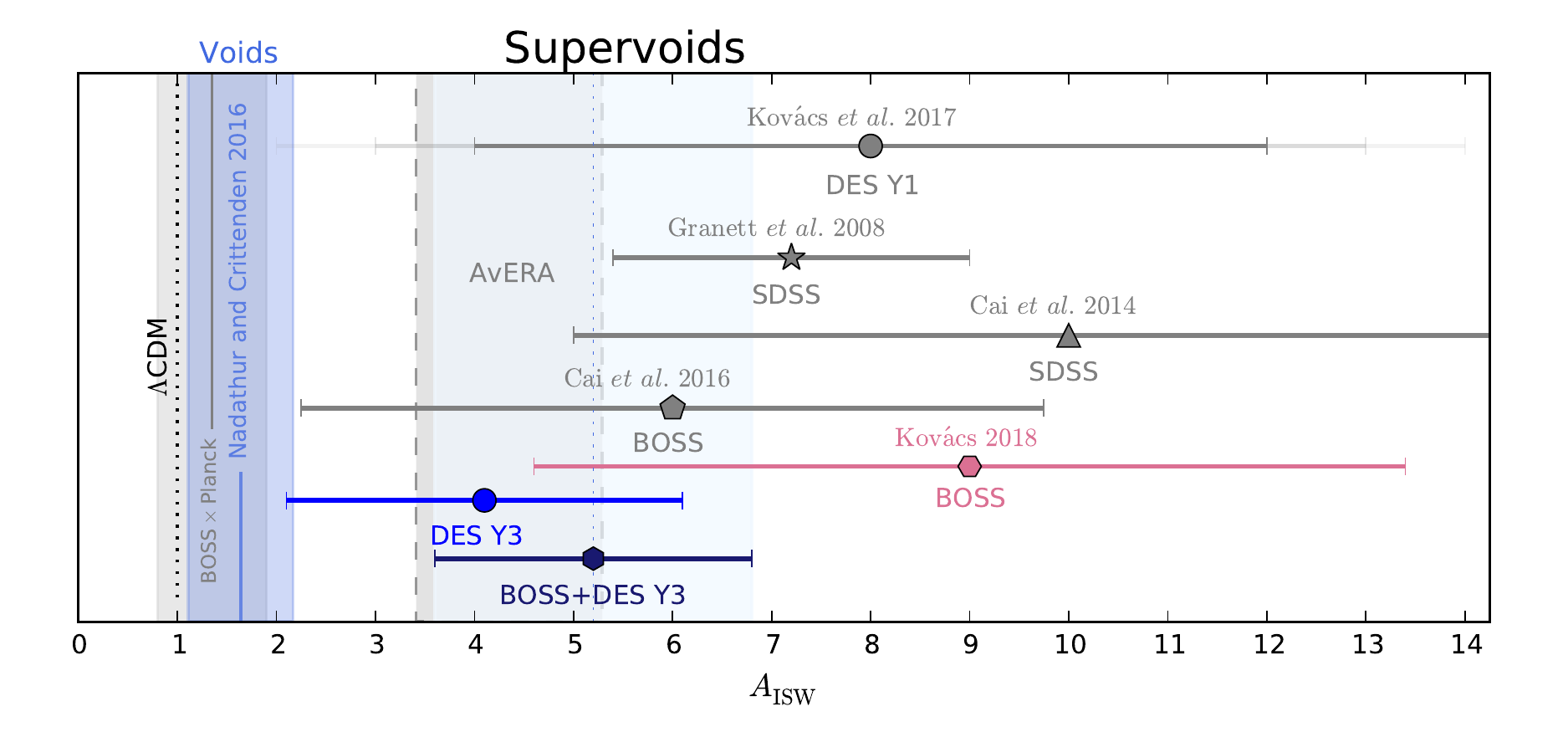}
\caption{Summary of our DES Y3 and combined DES+BOSS results, and a comparison to other supervoid-based estimates of the $A_{\rm ISW}$ amplitude (with $1\sigma$ errors shown). We assumed $\Delta T_{f}=-1.33~\mu K$, estimated by Nadathur et al. (2012), for the measurement by Granett et al. who used SDSS data. See Cai et al. (2014, 2016) for their simulation estimates of the $\Lambda$CDM signal. The circle marker with different error bars corresponds to separate Y1 DES void, DES supercluster, and combined DES Y1 constraints. The rose colored symbol marks the BOSS DR12 result by Kov\'acs (2018) that we use for combination in this study. The dotted vertical black line marks the Jubilee $\Lambda$CDM expectation with $A_{\rm ISW}=1$. The gray band shows $1\sigma$ constraints based on angular cross-correlations using  {\it Planck} and BOSS data, while the light blue band corresponds to $1\sigma$ constraints by Nadathur $\&$ Crittenden (2016). The AvERA simulation's higher expected amplitude, restricted to multipoles $10<\ell<100$, is illustrated with the vertical band in between gray dashed lines.}
\end{center}
\end{figure*}

The combination with independent BOSS data reveals an excess ISW signal of supervoids with $A_{\rm ISW}\approx5.2\pm1.6$ amplitude, corresponding to a $3.3\sigma$ observation (see Table 2). The tension with $\Lambda$CDM predictions is equivalent to $2.6\sigma$. This moderately significant observation of excess ISW-like signals of supervoids is consistent with several previous estimates based, at least in part, on merged voids \citep{GranettEtAl2008,CaiEtAl2014,Cai2016,Kovacs2016,Kovacs2017}, but inconsistent with presumably more optimal measurements that use full cross-correlation or a more detailed decomposition of the cosmic web as an estimator \citep[see e.g.][]{PlanckISW2015, NadathurCrittenden2016, Stolzner2017}.

The origin of this excess is of course not necessarily cosmological but at least the signal has now been proven to be robust. We also know that it shows no CMB color dependence which excludes some possibilities. A residual contamination, coming from unresolved extragalactic point sources, may still be blamed though \citep[see e.g.][]{Millea2012}, since dust from galaxies at all redshifts contributes to the CMB temperature fluctuations, which, in turn, would result in a positive correlation between CMB temperatures and galaxy density \citep[see e.g.][]{ho}. However, \cite{Hernandez2013} reported that realistic contaminations of this kind leave a different imprint than what has been observed by \cite{GranettEtAl2008} thus this possibility seems unlikely.

\begin{table}
\centering
\caption{Summary of our main results using DES and BOSS supervoids, including a measure of tension between observations and the Jubilee $\Lambda CDM$ estimates.}
\begin{tabular}{@{}cccc}
\hline
\hline
Cross-correlated data sets & $A_{\rm ISW}$ & S/N  & Tension \\
\hline
DES Y3 $\times$ Planck & $4.1\pm2.0$ & 2.1 & $1.6\sigma$\\
DES Y3+BOSS DR12 $\times$ Planck & $5.2\pm1.6$  & 3.3 & $2.6\sigma$\\
\hline
\hline
\end{tabular}
\label{table:merging}
\end{table}

In the theory ground, it is typically assumed that neither modifications of the concordance model, given other precise constraints \citep{Nadathur2012}, nor e.g. simple modified gravity scenarios seem to alleviate the ISW tension \citep[e.g.][]{CaiEtAl2014}. However, it is interesting to see that the amount of the excess signal is close to what is expected in the AvERA (Average Expansion Rate Approximation) cosmology. \cite{Racz2016} calculate the expansion rate of
local mini-universes and average the volume increment spatially to get the global scale factor increment for an otherwise normal $N$-body simulation. Recently, \cite{Beck2018} reported an enhanced $3.4 < A_{\rm ISW} < 5.3$ amplitude using ISW auto-correlation functions for multipoles $10<\ell<100$ in the AvERA model. This excess is sourced by a characteristically higher growth factor derivative at $z\lsim1.5$ where these supervoids may reprocess the CMB light. In general, modified gravity theories with alternative growth rates might provide some ground to discuss such excess signals, related especially to spatial perturbations in dark energy that are expected to mainly alter large-scale physics and their unique ISW effect is their main hope to be uncovered \citep[see e.g.][]{WellerLewis2003,Bean2004,HuScranton2004,Mota2008,dePutter2010}. 

All things considered, the observational evidence for an excess ISW signal of supervoids is still not fully conclusive but certainly warrants further studies as it practically remains unexplained. There are, however, other anomalies in cosmology that may be related to the ISW mystery. 

Firstly, our results provide a framework to re-think the qualitatively similar case of the CMB Cold Spot \citep[see e.g.][]{CruzEtal2004, FinelliEtal2014, SzapudiEtAl2014, KovacsJGB2015, Naidoo2016, Naidoo2017}. With a significant enhancement of the density-temperature correlation at large scales, the Cold Spot and the Eridanus supervoid can plausibly be related via this unexpected excess signal. Naively, the recent findings by \cite{Mackenzie2017}, based on a pencil beam-like galaxy survey centered on the Cold Spot, suggest that the Eridanus supervoid is not special and thus we should expect similar cold spots elsewhere in the sky where they see similar depression in pencil beam-like statistics. However, this approach ignores the role of the environment of these supervoids, or in other words fails to map the full extent of these large structures. More recently, \cite{Courtois2017} showed that a significant basin of repulsion is located in the proximate direction toward the Cold Spot, using the wide-area three-dimensional gravitational velocity field within $z\approx0.1$. With wide-angle DES \redmagic\ photo-$z$ data in the Cold Spot area, we will map this area in greater details (Kov\'acs et al. in prep.) to see the full extent of the Eridanus supervoid. We note that the biggest supervoid identified in the present Y3 analysis is of radius $R_{v}\approx246~\mpc$ and of central galaxy underdensity $\delta_g^{c}\approx-0.6$ (equivalent to a matter underdensity of $\delta_m^{c}\approx-0.3$ with the simple DES \redmagic\ linear galaxy bias $b_{g}\approx2.0$). While these void parameters are comparable to those of the Eridanus supervoid, the redshift is higher in this case with $z\approx0.5$ (corresponding to a smaller angular size) thus the expected ISW imprint is less strong and harder to clearly detect than that of a supervoid at $z\approx0.1$.

Secondly, these new findings raise the possibility that the moderate tensions between cosmological parameters determined from the late Universe and those from CMB anisotropies may arise because both the local cosmic web \emph{and} the CMB maps are in fact problematic. Speculatively, supervoids and inhomogeneities may be more influential than expected. This can contribute to biases in the determination of the Hubble constant \citep[see e.g.][]{Bernal2016}, and at the same time imprint unexpected secondary anisotropies in the CMB that can, in principle, alter several cosmological observables and lead to anomalies \citep{Schwarz2015}.

In summary, we argue that smaller catalogues of the largest voids may be more informative about dark energy than presumably optimal techniques. We observe possible problems at the largest scales and averaging may wash out the interesting new features. In the Dark Energy Survey Collaboration, we aim to continue this research along slightly different lines, including measurements of the imprint of these supervoids in CMB lensing convergence maps, and extensions of the measurements to supercluster samples. In the near future, we believe that, beyond a better understanding of the methodologies and possible re-analyses, new cosmic web decomposition data from experiments like the Dark Energy Spectroscopic Instrument (DESI) \citep{DESI} and the Euclid mission \citep{euclid} will further constrain the ISW signal.

\section*{Acknowledgments}
This paper has gone through \emph{internal review} by the DES collaboration. It has been assigned DES paper ID DES-2018-0343 and IFT preprint number IFT-UAM/CSIC-19-005.

Funding for the DES Projects has been provided by the U.S. Department of Energy, the U.S. National Science Foundation, the Ministry of Science and Education of Spain, the Science and Technology Facilities Council of the United Kingdom, the Higher Education Funding Council for England, the National Center for Supercomputing Applications at the University of Illinois at Urbana-Champaign, the Kavli Institute of Cosmological Physics at the University of Chicago, the Center for Cosmology and Astro-Particle Physics at the Ohio State University, the Mitchell Institute for Fundamental Physics and Astronomy at Texas A\&M University, Financiadora de Estudos e Projetos, 
Funda{\c c}{\~a}o Carlos Chagas Filho de Amparo {\`a} Pesquisa do Estado do Rio de Janeiro, Conselho Nacional de Desenvolvimento Cient{\'i}fico e Tecnol{\'o}gico and the Minist{\'e}rio da Ci{\^e}ncia, Tecnologia e Inova{\c c}{\~a}o, the Deutsche Forschungsgemeinschaft and the Collaborating Institutions in the Dark Energy Survey. 

The Collaborating Institutions are Argonne National Laboratory, the University of California at Santa Cruz, the University of Cambridge, Centro de Investigaciones Energ{\'e}ticas, Medioambientales y Tecnol{\'o}gicas-Madrid, the University of Chicago, University College London, the DES-Brazil Consortium, the University of Edinburgh, the Eidgen{\"o}ssische Technische Hochschule (ETH) Z{\"u}rich, 
Fermi National Accelerator Laboratory, the University of Illinois at Urbana-Champaign, the Institut de Ci{\`e}ncies de l'Espai (IEEC/CSIC), 
the Institut de F{\'i}sica d'Altes Energies, Lawrence Berkeley National Laboratory, the Ludwig-Maximilians Universit{\"a}t M{\"u}nchen and the associated Excellence Cluster Universe, the University of Michigan, the National Optical Astronomy Observatory, the University of Nottingham, The Ohio State University, the University of Pennsylvania, the University of Portsmouth, SLAC National Accelerator Laboratory, Stanford University, the University of Sussex, Texas A\&M University, and the OzDES Membership Consortium.

Based in part on observations at Cerro Tololo Inter-American Observatory, National Optical Astronomy Observatory, which is operated by the Association of Universities for Research in Astronomy (AURA) under a cooperative agreement with the National Science Foundation.

The DES data management system is supported by the National Science Foundation under Grant Numbers AST-1138766 and AST-1536171. The DES participants from Spanish institutions are partially supported by MINECO under grants AYA2015-71825, ESP2015-66861, FPA2015-68048, SEV-2016-0588, SEV-2016-0597, and MDM-2015-0509, 
some of which include ERDF funds from the European Union. IFAE is partially funded by the CERCA program of the Generalitat de Catalunya. AK has also been supported by a Juan de la Cierva fellowship from MINECO.
Research leading to these results has received funding from the European Research
Council under the European Union's Seventh Framework Program (FP7/2007-2013) 
including ERC grant agreements 240672, 291329, 306478, and 615929.
We acknowledge support from the Australian Research Council Centre of Excellence for All-sky Astrophysics (CAASTRO), through project number CE110001020, and the Brazilian Instituto Nacional de Ci\^enciae Tecnologia (INCT) e-Universe (CNPq grant 465376/2014-2).

This manuscript has been authored by Fermi Research Alliance, LLC under Contract No. DE-AC02-07CH11359 with the U.S. Department of Energy, Office of Science, Office of High Energy Physics. The United States Government retains and the publisher, by accepting the article for publication, acknowledges that the United States Government retains a non-exclusive, paid-up, irrevocable, world-wide license to publish or reproduce the published form of this manuscript, or allow others to do so, for United States Government purposes.

The authors thank the Jubilee team for providing their LRG mock data and ISW maps based on their N-body simulation that was performed on the Juropa supercomputer of the J\"ulich Supercomputing Centre (JSC).

This work has made use of public data from the SDSS-III collaboration. Funding for SDSS-III has been provided by the Alfred P. Sloan Foundation, the Participating Institutions, the National Science Foundation, and the U.S. Department of Energy Office of Science. The SDSS-III website is http://www.sdss3.org/. 

\bibliographystyle{mn2e}
\bibliography{refs}

\section*{Affiliations}
$^{1}$ Institut de F\'{\i}sica d'Altes Energies (IFAE), The Barcelona Institute of Science and Technology, Campus UAB, 08193 Bellaterra (Barcelona), Spain\\
$^{2}$ Instituto de Astrof\'{\i}sica de Canarias (IAC), Calle V\'{\i}a L\'{a}ctea, E-38200, La Laguna, Tenerife, Spain;
Departamento de Astrof\'{\i}sica, Universidad de La Laguna (ULL), E-38206, La Laguna, Tenerife, Spain\\
$^{3}$ Department of Physics and Astronomy, University of Pennsylvania, Philadelphia, PA 19104, USA\\
$^{4}$ Instituto de F\'isica Te\'orica IFT-UAM/CSIC, Universidad Aut\'onoma de Madrid, Cantoblanco 28049 Madrid, Spain\\
$^{5}$ Jodrell  Bank  Center  for  Astrophysics,  School  of  Physics  and  Astronomy,
University  of  Manchester,  Oxford  Road,  Manchester,  M13  9PL,  UK\\
$^{6}$ Universit\"ats-Sternwarte, Fakult\"at f\"ur Physik, Ludwig-Maximilians Universit\"at M\"unchen, Scheinerstr. 1, 81679 M\"unchen, Germany\\
$^{7}$ Department of Astronomy/Steward Observatory, 933 North Cherry Avenue, Tucson, AZ 85721-0065, USA\\
$^{8}$ Institute of Cosmology \& Gravitation, University of Portsmouth, Portsmouth, PO1 3FX, UK\\
$^{9}$ Cerro Tololo Inter-American Observatory, National Optical Astronomy Observatory, Casilla 603, La Serena, Chile\\
$^{10}$ Department of Physics \& Astronomy, University College London, Gower Street, London, WC1E 6BT, UK\\
$^{11}$ Fermi National Accelerator Laboratory, P. O. Box 500, Batavia, IL 60510, USA\\
$^{12}$ Kavli Institute for Particle Astrophysics \& Cosmology, P. O. Box 2450, Stanford University, Stanford, CA 94305, USA\\
$^{13}$ SLAC National Accelerator Laboratory, Menlo Park, CA 94025, USA\\
$^{14}$ CNRS, UMR 7095, Institut d'Astrophysique de Paris, F-75014, Paris, France\\
$^{15}$ Sorbonne Universit\'es, UPMC Univ Paris 06, UMR 7095, Institut d'Astrophysique de Paris, F-75014, Paris, France\\
$^{16}$ Centro de Investigaciones Energ\'eticas, Medioambientales y Tecnol\'ogicas (CIEMAT), Madrid, Spain\\
$^{17}$ Laborat\'orio Interinstitucional de e-Astronomia - LIneA, Rua Gal. Jos\'e Cristino 77, Rio de Janeiro, RJ - 20921-400, Brazil\\
$^{18}$ Department of Astronomy, University of Illinois, 1002 W. Green Street, Urbana, IL 61801, USA\\
$^{19}$ National Center for Supercomputing Applications, 1205 West Clark St., Urbana, IL 61801, USA\\
$^{20}$ Physics Department, 2320 Chamberlin Hall, University of Wisconsin-Madison, 1150 University Avenue Madison, WI  53706-1390\\
$^{21}$ Institut de Ci\`encies de l'Espai, IEEC-CSIC, Campus UAB, Carrer de Can Magrans, s/n,  08193 Bellaterra, Barcelona, Spain\\
$^{22}$ Observat\'orio Nacional, Rua Gal. Jos\'e Cristino 77, Rio de Janeiro, RJ - 20921-400, Brazil\\
$^{23}$ George P. and Cynthia Woods Mitchell Institute for Fundamental Physics and Astronomy, and Department of Physics and Astronomy, Texas A\&M University, College Station, TX 77843,  USA\\
$^{24}$ Department of Physics, IIT Hyderabad, Kandi, Telangana 502285, India\\
$^{25}$ Department of Physics, University of Michigan, Ann Arbor, MI 48109, USA\\
$^{26}$ Department of Physics, ETH Zurich, Wolfgang-Pauli-Strasse 16, CH-8093 Zurich, Switzerland\\
$^{27}$ Santa Cruz Institute for Particle Physics, Santa Cruz, CA 95064, USA\\
$^{28}$ Center for Cosmology and Astro-Particle Physics, The Ohio State University, Columbus, OH 43210, USA\\
$^{29}$ Department of Physics, The Ohio State University, Columbus, OH 43210, USA\\
$^{30}$ Max Planck Institute for Extraterrestrial Physics, Giessenbachstrasse, 85748 Garching, Germany\\
$^{31}$ Harvard-Smithsonian Center for Astrophysics, Cambridge, MA 02138, USA\\
$^{32}$ Australian Astronomical Observatory, North Ryde, NSW 2113, Australia\\
$^{33}$ Departamento de F\'isica Matem\'atica, Instituto de F\'isica, Universidade de S\~ao Paulo, CP 66318, S\~ao Paulo, SP, 05314-970, Brazil\\
$^{34}$ Department of Astrophysical Sciences, Princeton University, Peyton Hall, Princeton, NJ 08544, USA\\
$^{35}$ Department of Astronomy, University of Michigan, Ann Arbor, MI 48109, USA\\
$^{36}$ Instituci\'o Catalana de Recerca i Estudis Avan\c{c}ats, E-08010 Barcelona, Spain\\
$^{37}$ Excellence Cluster Universe, Boltzmannstr.\ 2, 85748 Garching, Germany\\
$^{38}$ Jet Propulsion Laboratory, California Institute of Technology, 4800 Oak Grove Dr., Pasadena, CA 91109, USA\\
$^{39}$ Department of Physics and Astronomy, Pevensey Building, University of Sussex, Brighton, BN1 9QH, UK\\
$^{40}$ School of Physics and Astronomy, University of Southampton,  Southampton, SO17 1BJ, UK\\
$^{41}$ Brandeis University, Physics Department, 415 South Street, Waltham MA 02453\\
$^{42}$ Instituto de F\'isica Gleb Wataghin, Universidade Estadual de Campinas, 13083-859, Campinas, SP, Brazil\\
$^{43}$ Computer Science and Mathematics Division, Oak Ridge National Laboratory, Oak Ridge, TN 37831\\
$^{44}$ Argonne National Laboratory, 9700 South Cass Avenue, Lemont, IL 60439, USA

\end{document}